# Local Measurements of the Superconducting Pairing Symmetry in $Cu_xBi_2Se_3$

Niv Levy[1,2], Tong Zhang[1,2], Jeonghoon Ha[1,2,3], Fred Sharifi[1], A. Alec Talin[1], Young Kuk[3], and Joseph A. Stroscio[1]

[1] Center for Nanoscale Science and Technology, NIST, Gaithersburg, MD 20899, USA
[2] Maryland NanoCenter, University of Maryland, College Park, MD 20742, USA
[3] Department of Physics and Astronomy, Seoul National University, Seoul, 151-747, Korea

Topological superconductors represent a newly predicted phase of matter that is topologically distinct from conventional superconducting condensates of Cooper pairs. As a manifestation of their topological character, topological superconductors support solid-state realizations of Majorana fermions at their boundaries. The recently discovered superconductor $Cu_xBi_2Se_3$ has been theoretically proposed as an odd-parity superconductor in the time-reversal-invariant topological superconductor class and point-contact spectroscopy measurements have reported the observation of zero-bias conductance peaks corresponding to Majorana states in this material. Here we report scanning tunneling spectroscopy (STS) measurements of the superconducting energy gap in $Cu_xBi_2Se_3$ as a function of spatial position and applied magnetic field. The tunneling spectrum shows that the density of states at the Fermi level is fully gapped without any in-gap states. The spectrum is well described by the Bardeen-Cooper-Schrieffer (BCS) theory with a momentum independent order parameter, which suggests that $Cu_{0.2}Bi_2Se_3$ is a classical $s$-wave superconductor contrary to previous expectations and measurements.

Recently the discovery of topological insulators has led to a new classification paradigm of quantum states of matter based on band topology [1,2]. Topological insulators are time-reversal-invariant materials topologically distinct from conventional insulators. In analogy to the notion of topological insulators, a new class of topological superconductors has also been predicted. These are topologically distinct from normal Cooper pair condensates and can host delocalized Andreev surface states [1,2]. Such surface states are a condensed-matter realization of Majorana fermions, which have potential uses in topologically protected quantum computation [3]. Motivated by the possible realization of Majorana fermions, the search for topological superconductors has attracted significant attention. In the case of time-reversal-invariant centrosymmetric materials, Fu and Berg [4] have found a set of criteria sufficient for establishing whether a material is a topological superconductor. These criteria stipulate that a material is a topological superconductor if it is an odd-parity, fully gapped superconductor and its Fermi surface encloses an odd number of time-reversal-invariant momenta in the Brillouin zone. $Cu_xBi_2Se_3$, which is transformed from the topological insulator $Bi_2Se_3$ into a superconductor by intercalating Cu between the weakly bonded Se layers, has been considered a promising candidate; superconducting $Cu_xBi_2Se_3$ occurs with a Cu concentration of $0.1 < x < 0.6$ [5–9]. $Cu_xBi_2Se_3$ potentially fits the criteria set forth by Fu and Berg because the Fermi surface encloses the time-reversal-invariant momentum state at **k**=0 and has the possibility of odd-parity pairing symmetry due to strong spin-orbit coupling [4,10]. The pairing symmetries of $Cu_xBi_2Se_3$ are theoretically classified based on its crystal point group symmetry [4,10,11]. Out of the four possibilities found in reference [4], three are odd-parity, unconventional paired states. The observation of a zero-bias conductance peak in recent point-contact spectroscopy measurements [9,12,13] has supported the possibility of unconventional pairing symmetry, and was interpreted [9–11,13] as a signature of the topological surface states associated with nontrivial topological superconductivity.

Tunneling spectroscopy has played an important role in determining the gap function and pairing symmetries of superconductors since the 1960s [14]. A tunneling spectrum with zero-bias anomalies has been attributed to nodes in the order parameter of *d*-wave high-temperature superconductors [15], and to possible *p*-wave pairing in $Sr_2RuO_4$ [16]. In this report, we present the first measurements of the local tunneling spectrum of $Cu_xBi_2Se_3$ and its response to an applied magnetic field.

Pristine $Bi_2Se_3$ has a layered structure consisting of Se-Bi-Se-Bi-Se quintuple layers (Fig. 1(a)), where the coupling between adjacent quintuple layers is weak. This makes the crystal cleavable and results in good quality Se terminated (111) planes (Fig. 1(b)). The <111> rhombohedral direction is denoted as the c-axis. While the pristine material is a topological insulator with a bandgap of 0.3 eV [1], $Cu_xBi_2Se_3$ is a superconductor with a transition temperature of 2.2 K to 3.8 K for x between 0.1 and 0.6 [5–9]. A comparison of structural and transport measurements have determined that intercalated copper is responsible for the superconducting phase as opposed to copper which is substituted in the Bi lattice [5].

We prepared $Cu_{0.2}Bi_2Se_3$ by copper intercalation of a $Bi_2Se_3$ crystal, following the methods outlined by Kriener *et al*. [8]. The $Bi_2Se_3$ crystal was prepared by melting a 1:3 mixture of 5N purity Bi and Se shot, respectively, under $H_2$-Ar forming gas flow at 740 °C, followed by cooling to 550 °C over 48 h and an 80 h long anneal at 550 °C. The resulting crystal was found to cleave easily and to be of high quality through X-ray diffraction and scanning tunneling microscopy (STM) measurements. A sample of approximate size $3.5 \times 3.8 \times 0.4$ mm$^3$ was cut from a pristine crystal and electrochemically intercalated with Cu by suspension in a saturated CuI solution and galvanostatically applying a current of 11.5 µA for 45 h, using a Cu wire as the anode. The average copper fraction is $0.2 \pm 0.03$ [17], as determined from the weight difference before and after the intercalation. The intercalated crystal was annealed at 535 °C for 2 h and then quenched in water. Resistance versus temperature measurements show a superconducting transition at $3.65 \pm 0.05$ K, and tunneling spectroscopy measurements (see Fig. 4(a)) show a critical field of $B_{C2} = 1.65 \pm 0.05$ T [17] in agreement with previous measurements [5,6], indicating that the active component of our sample is the same as that studied by Kriener *et al*. [6]. The sample was cleaved in ultra-high vacuum at $4 \times 10^{-9}$ Pa to expose the (111) Se plane and then transferred *in-situ* to a custom cryogenic STM system [18] prior to cooling down to 15 mK. The first derivative of the tunneling current, $I$, with respect to the tunneling sample bias, $V_B$, was obtained using lock-in detection with a modulation frequency of ≈500 Hz, and with modulation amplitudes (root-mean-square) $V_{mod}$ from 5 µV to 50 µV depending on the spectral range of interest. An effective energy resolution of 280 mK in the tunneling measurements was determined by fitting the tunneling spectrum from an *in situ* grown Al sample [19].

Large scale topography (Fig. 1(c)) of the cleaved $Cu_{0.2}Bi_2Se_3$ crystal shows a surface with atomically flat terraces hundreds of nanometers wide. The terraces (Fig. 1(d)) are decorated by small clusters with different density, presumably Cu, as seen in the previous STM studies [5]. The Cu clusters are also seen to decorate grain boundaries (Fig. 1(e)). The number of grain boundaries is large compared with what is seen in the $Bi_2Se_3$ samples prior to the intercalation, which likely stems from the quench employed during sample processing. The granularity may play a role in the emergence of superconductivity in $Cu_xBi_2Se_3$ [20]. A small scale view of a terrace area (Fig. 1(f)) shows a clear $Bi_2Se_3$ (111) hexagonal atomic lattice with a lattice constant of 0.41 nm, without any apparent distortion.

The local order parameter of the superconducting state was characterized by scanning tunneling spectroscopy (STS). The differential tunneling conductance between the probe tip and sample, $dI/dV$, is used to determine the superconducting energy gap as a function of spatial position. Figure 2(a) shows the conductance spectrum covering a voltage range of ±30 mV, which displays an asymmetric varying background and superconducting gap at zero bias. Focusing on the low energy region near zero bias, the single point spectrum in Fig. 2(b), obtained from the region in Fig. 1(e), clearly shows a fully gapped superconducting state with pronounced coherence peaks. We have fitted the spectra to the Maki extension of the BCS theory, using the known effective temperature [19] and an adjustable depairing parameter $\zeta$ that characterizes the Cooper pair lifetime [21]. The Maki theory with $\zeta=0$ reduces to the BCS density of states. The fit to the spectrum in Fig. 2(b) yields an energy gap $\Delta = 0.399 \pm 0.001$ meV and $\zeta = 0.0092 \pm 0.0002$ [22]. The spectrum fits the BCS based model well and is fully gapped with zero differential conductance within $2\sigma$ of the noise level over 0.54 mV about zero bias. This strongly suggests that $Cu_xBi_2Se_3$ is an *s*-wave superconductor. Other possible pairing symmetries related to topologically nontrivial odd-parity pairing discussed in relation to topological superconductors produce tunneling spectra that are either not fully gapped or have a finite density of states at zero bias due to surface states [10,11], neither of which is observed in the spectrum in Fig. 2(b).

However, we found that in some cases zero bias peaks can be observed in low impedance measurements (Fig. 2(c)), specifically when the tip also became contaminated and possibly superconducting after crashing into the $Cu_xBi_2Se_3$ sample, forming a superconductor-insulator-superconductor (SIS) junction. In this case the zero bias peaks at low impedances would be due to Josephson tunneling in the SIS tunnel junction [23,24] and not due to Majorana states. In such cases, an asymmetric superconducting tunnel junction can form with the probe tip having a smaller gap, $\Delta_{Tip}$, due to the small size of the superconducting microcrystal on the probe tip [24]. The resulting asymmetric junction is characterized by a series of peaks corresponding to multiple Andreev reflections at energies of $\pm\Delta_{Tip}$, $\pm\Delta_{Sample}$, $\pm(\Delta_{Sample} - \Delta_{Tip})$, and $\pm(\Delta_{Sample} + \Delta_{Tip})/(2m+1)$ where $m$ is an integer [24–26]. The strength of the peaks is dependent on junction resistance and magnitude of $\Delta_{Tip}$ compared to $\Delta_{Sample}$ [24]. A fit of the spectra at high impedance (1 MΩ in Fig. 2(c)) to the SIS model yields $\Delta_{Sample} = 0.44$ meV and $\Delta_{Tip} = 0.16$ meV. Experimentally, peaks are observed in Fig.2 (c) at $\pm(\Delta_{Sample} + \Delta_{Tip})$ for all junction resistances, and at low junction resistances peaks develop at $\pm(\Delta_{Sample} - \Delta_{Tip})$ and at zero bias. The fit to the SIS model suggests that the zero bias peak observed at low junction resistance results from the Josephson supercurrent between the two superconductors. In contrast, tunneling into $Cu_xBi_2Se_3$ with a normal probe tip produces no zero bias peaks, as shown in Fig. 2(a), and as seen in measurements on another cleaved sample with a new non-superconducting probe tip even at junction impedances down to 3 kΩ (Fig. 2(d)).

We have found the $Cu_xBi_2Se_3$ sample to display large inhomogeneity in the superconducting state (Fig. 3). Within superconducting grains (Fig. 3(a)) the gap varies very little even as we traverse terraces with grain boundaries and Cu clusters as seen in the spectral profile in Fig. 3(b) where the gap averaged over the 135 spectra in Fig. 3(b) is $0.41 \pm 0.03$ meV. However, there are also regions where non-superconducting terraces border superconducting areas. Figure 3(c) shows one of these areas where the upper left region in the image is superconducting and the terrace on the lower right is normal. A *dI/dV* line scan from the indicated region in Fig. 3(c) is shown in Fig. 3(d) where the gap is seen to decay on the scale of ≈30 nm as we go from the area in the top left to bottom right. The decay in the gap across the superconducting – normal boundary is consistent with the coherence length found in macroscopic measurements of $Cu_xBi_2Se_3$ [6].

To study the possible emergence of the Majorana state in the vortex core of the type II superconductor $Cu_{0.2}Bi_2Se_3$ [2], we have also carried out field dependent spectroscopy measurements. With increasing magnetic field, the superconducting gap was found to decrease and disappear above 1.7 T (Fig. 4(a)). A BCS fit to the spectra yields the field dependence of the gap shown in Fig. 4(b), with a critical field of $B_{C2} = 1.65 \pm 0.05$ T [19]. This is in good agreement with previous bulk measurements performed by other groups [5,6]. A spatial mapping of the differential conductivity at zero bias reveals the emergence of vortices, which increase in density with increasing field (Fig. 4(d)-(f)). Inside the vortex core (Fig. 4(c)) we observe the absence of the superconducting gap over the length scale of ≈20 nm. We did not observe any zero bias states related to the Majorana states inside the cores [20]. This may be related to the relatively high doping level in $Cu_{0.2}Bi_2Se_3$ [27], which may inhibit the formation of the Majorana modes in the vortex cores for $Cu_{0.2}Bi_2Se_3$ with the field applied along the *c*-axis according to recent theory [28].

The spectra presented here lead to a markedly different conclusion than that reached by point contact spectroscopy studies, where zero bias conductance peaks were observed and interpreted in terms of nontrivial topological superconductivity [9,12,13]. We have observed a superconducting gap consistent with an *s*-wave pairing and did not observe any zero bias anomalies with a normal tip. To reconcile these differences we can examine the differences in the measurements and possible differences between the samples. In our tunneling measurements we varied the junction resistance from G$\Omega$s to k$\Omega$s, which corresponds to measurements from the vacuum tunneling range to point contact. However in our point contact measurements the junction is very controlled through a few atoms without damaging the crystal surface. In macroscopic point contact one has to question what is the area and structure of the contact, and how is the crystal structure modified at the contact point. It is possible that Majorana modes may exist on other crystal faces that point contact measurements access, which are not present on the (111) surface probed in the tunneling measurements. Furthermore, zero bias peaks can sometimes be observed in point contact spectroscopy of *s*-wave superconductors [29,30], and therefore their observation may be an indication of unconventional parity, but not a sufficient one. An additional consideration is whether the measurements were obtained on similar samples having the same superconducting phase. The original point contact study [9] was on material with a higher Cu concentration, x=0.3 versus x=0.2 in the present study. It is therefore possible that the novel pairing phase develops at a higher Cu concentration. However, our critical field measurements match previous bulk measurements and more recent point contact spectroscopy measurements reported zero bias states with x=0.25 Cu doping [13], close to the present study.

In summary, we presented the first low temperature tunneling spectroscopy measurements of $Cu_xBi_2Se_3$. Based on the tunneling spectra we conclude that most likely $Cu_xBi_2Se_3$ is a *s*-wave superconductor but that future tunneling spectroscopy on higher doped material will be useful to examine the full superconducting phase range of this new material.

**Acknowledgements**: We thank Liang Fu, Timothy Hsieh, and Yoichi Ando for valuable discussions, and Steve Blankenship and Alan Band for technical assistance, and Nikolai Klimov for assistance in measuring transport properties. N.L., T.Z., and J.H acknowledges support under the Cooperative Research Agreement between the University of Maryland and the National Institute of Standards and Technology Center for Nanoscale Science and Technology, Award 70NANB10H193, through the University of Maryland. J.H. and Y.K. are partly supported by Korea Research Foundation through KRF-2010-00349.

**Figure Captions**:

Fig. 1. Structural characterization of the cleaved (111) surface of $Cu_{0.2}Bi_2Se_3$. (a) Structural model of $Cu_xBi_2Se_3$ (side view). The $p_z$ orbitals involved in the odd symmetry pairing models discussed in the text are indicated on the top and bottom Se layers of the top quintuple layer (red line). (b) Structural model of the top Se plane of $Cu_xBi_2Se_3$ (top view, Se atoms large orange, Bi atoms purple, unit cell green). (c) Large scale STM topography measurements of the cleaved (111) $Cu_{0.2}Bi_2Se_3$ surface showing large scale atomically flat terraces separated by multilayer steps. Image size is 1 μm x 1 μm. Tunneling parameters: $I$=10 pA, $V_B$=0.5 V. (d) Medium scale STM topography, 50 nm x 50 nm, showing cluster decoration and grain boundaries on the cleaved surface. Tunneling parameters: $I$=50 pA, $V_B$ =0.1 V. (e) Atomic resolution STM image, 10 nm x 10 nm, of the grain boundary in the blue boxed area in (d). Tunneling parameters: $I$=100 pA, $V_B$ =0.05 V. (f) Atomic resolution STM image of the top Se plane in the cleaved surface of $Cu_{0.2}Bi_2Se_3$ showing the atomic unit cell (white parallelogram). Image size is 4 nm x 4 nm. Tunneling parameters: $I$=20 pA, $V_B$ =-10 mV.

Fig. 2: Tunneling spectroscopy of $Cu_{0.2}Bi_2Se_3$. (a) Tunnel spectrum covering ±30 mV showing the background and superconducting gap near zero bias. Tunnel parameters: $I$=70 pA, $V_B$ =29 mV, $V_{mod}$=100 μV. (b) Low energy tunnel spectrum (red dots) showing a fully gapped $Cu_{0.2}Bi_2Se_3$ consistent with BCS $s$-wave pairing. A fit (blue line) to BCS Maki theory with $T_{eff}$ = 280 mK yields Δ=0.399±0.001 meV and ζ=0.0092±0.0002 [22]. Tunnel parameters: $I$=300 pA, $V_B$ =2 mV, $V_{mod}$=15 μV. (c) Tunnel spectrum of $Cu_{0.2}Bi_2Se_3$ with a probe tip crashed into the sample as a function of junction impedance. At high impedance the spectrum is well fit (red dashed line) by a BCS SIS model with the sample $Δ_{Sample}$=0.44± 0.01 meV and a probe tip gap $Δ_{Tip}$=0.16±0.01 meV, and $T_{eff}$ = 390±100 mK [22]. At lower junction impedance the spectrum develops a peak at zero bias due to Josephson tunneling. Tunnel parameters: $V_B$ =2 mV, $V_{mod}$ =10 μV, $I$=2 nA to 900 nA. (d) Tunneling spectrum on a second cleavage of the sample in (a) with a new probe tip showing no evidence of zero bias states at low impedance with a normal probe tip.

Fig. 3. Superconducting energy gap inhomogeniety in $Cu_{0.2}Bi_2Se_3$ . (a) STM topographic image, 50 nm x 50 nm, of a superconducting terrace region. The white line indicates where the *dI/dV* measurements were made in (b). (b) *dI/dV* spectra from the region in (a) showing a relatively uniform energy gap across the terraces, grain boundaries, and Cu clusters in (a). The average

gap from BSC fits is 0.41 ± 0.03 meV. Tunnel parameters: $I$=400 pA, $V_B$ =5 mV, $V_{mod}$ =20 µV. (c) STM topographic image, 160 nm x 160 nm, of a superconducting - normal interface region. The white line indicates where the *dI/dV* measurements were made in (d). (d) *dI/dV* spectral map along the line in (c) showing the gradual disappearance of the gap across the boundary. Each horizontal row is the *dI/dV* vs. $V_B$ spectra, where the *dI/dV* magnitude is given by the color intensity scale and the vertical axis is the distance along the line in (c). Tunnel parameters: $I$=200 pA, $V_B$ =5 mV, $V_{mod}$ =20 µV.

Fig. 4. Superconducting properties of $Cu_{0.2}Bi_2Se_3$ in an applied magnetic field. (a) Tunneling spectra as a function of an applied perpendicular magnetic field, B. The spectra show a transition to the normal state above $B$=1.7 T. The spectra are normalized to the spectra at high field ($B$=1.75-2 T). Tunnel parameters: $I$=80 pA, $V_B$ =10 mV, $V_{mod}$ =50 µV, $T_{eff}$=0.95 K. (b) The superconducting energy gap determined from fitting the spectra in (a) to the BCS density of states. The uncertainties given by one standard deviation in the gap energy determined from the fit to the BCS density of states are smaller than the symbol size [22]. (c) *dI/dV* spectral map showing the disappearance of the superconducting gap in the center of the vortex core in the top of the image in (f). Each vertical line is a *dI/dV* vs $V_B$ spectra, and the horizontal axis is the distance along the line in (f). The color scale is the *dI/dV* intensity from 0 to 30 nS. (d-f) Density of states maps at the Fermi level, *dI/dV* ($V_B$ =0 mV) / *dI/dV* ($V_B$ =-0.5 mV), 110 nm x 110 nm, showing the evolution of vortices (orange circles) as a function of applied magnetic field, (d) 0.5 T, (e) 0.75 T, and (f) 1.0 T. White line in (f) indicates path for the spectral map in (c). Tunnel parameters: $I$=200 pA, $V_B$ =5 mV, $V_{mod}$ =50 µV.

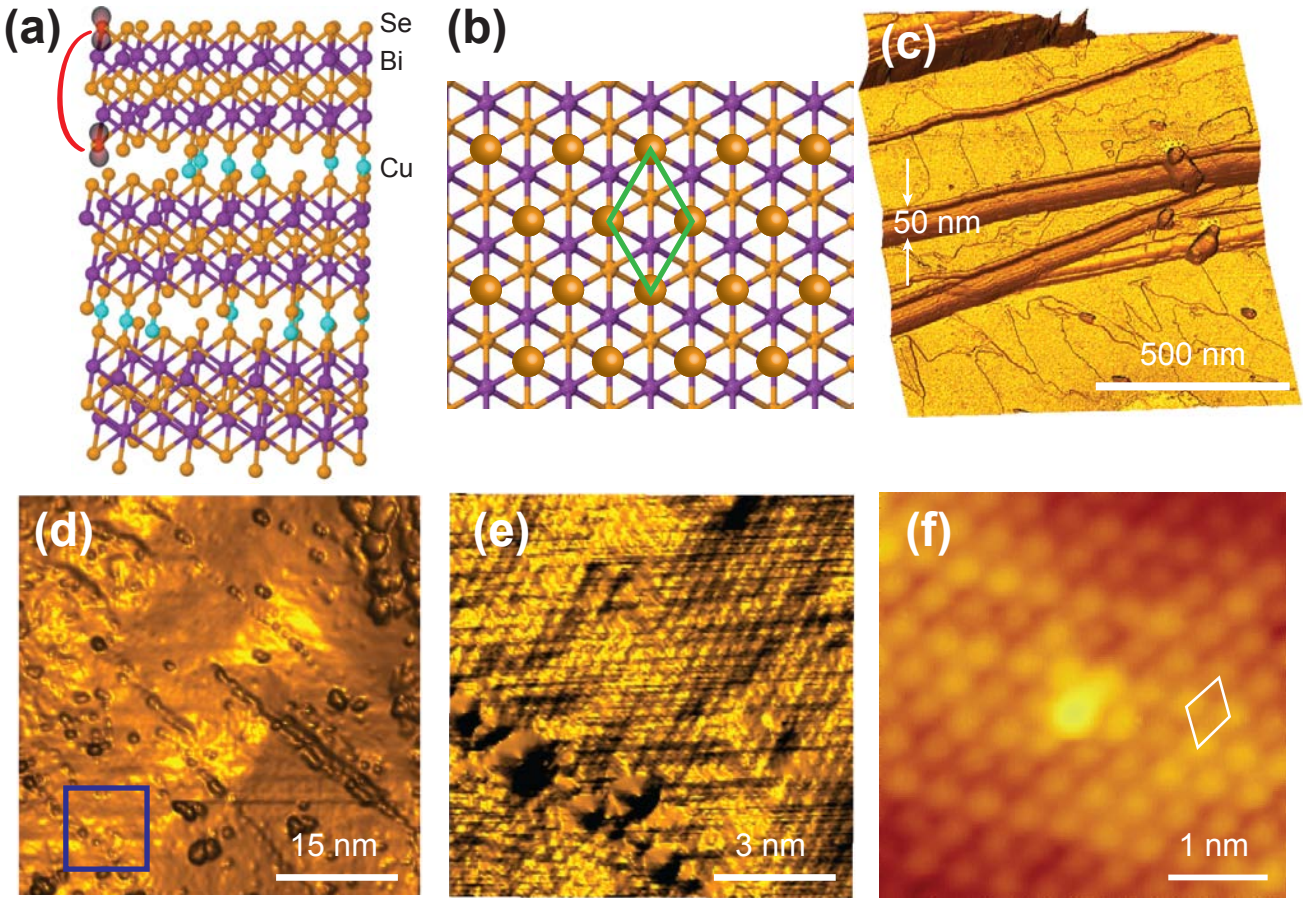

**Fig. 1**. Structural characterization of the cleaved (111) surface of $Cu_{0.2}Bi_2Se_3$. (**a**) Structural model of $Cu_xBi_2Se_3$ (side view). The $p_z$ orbitals involved in the odd symmetry pairing models discussed in the text are indicated on the top and bottom Se layers of the top quintuple layer (red line). (**b**) Structural model of the top Se plane of $Cu_xBi_2Se_3$ (top view, Se atoms large orange, Bi atoms purple, unit cell green). (**c**) Large scale STM topography measurements of the cleaved (111) $Cu_{0.2}Bi_2Se_3$ surface showing large scale atomically flat terraces separated by multilayer steps. Image size is 1 μm x 1 μm. Tunneling parameters: $I$=10 pA, $V_B$=0.5 V. (**d**) Medium scale STM topography, 50 nm x 50 nm, showing cluster decoration and grain boundaries on the cleaved surface. Tunneling parameters: $I$=50 pA, $V_B$=0.1 V. (**e**) Atomic resolution STM image, 10 nm x 10 nm, of the grain boundary in the blue boxed area in (d). Tunneling parameters: $I$=100 pA, $V_B$=0.05 V. (**f**) Atomic resolution STM image of the top Se plane in the cleaved surface of $Cu_{0.2}Bi_2Se_3$ showing the atomic unit cell (white parallelogram). Image size is 4 nm x 4 nm. Tunneling parameters: $I$=20 pA, $V_B$=-10 mV.

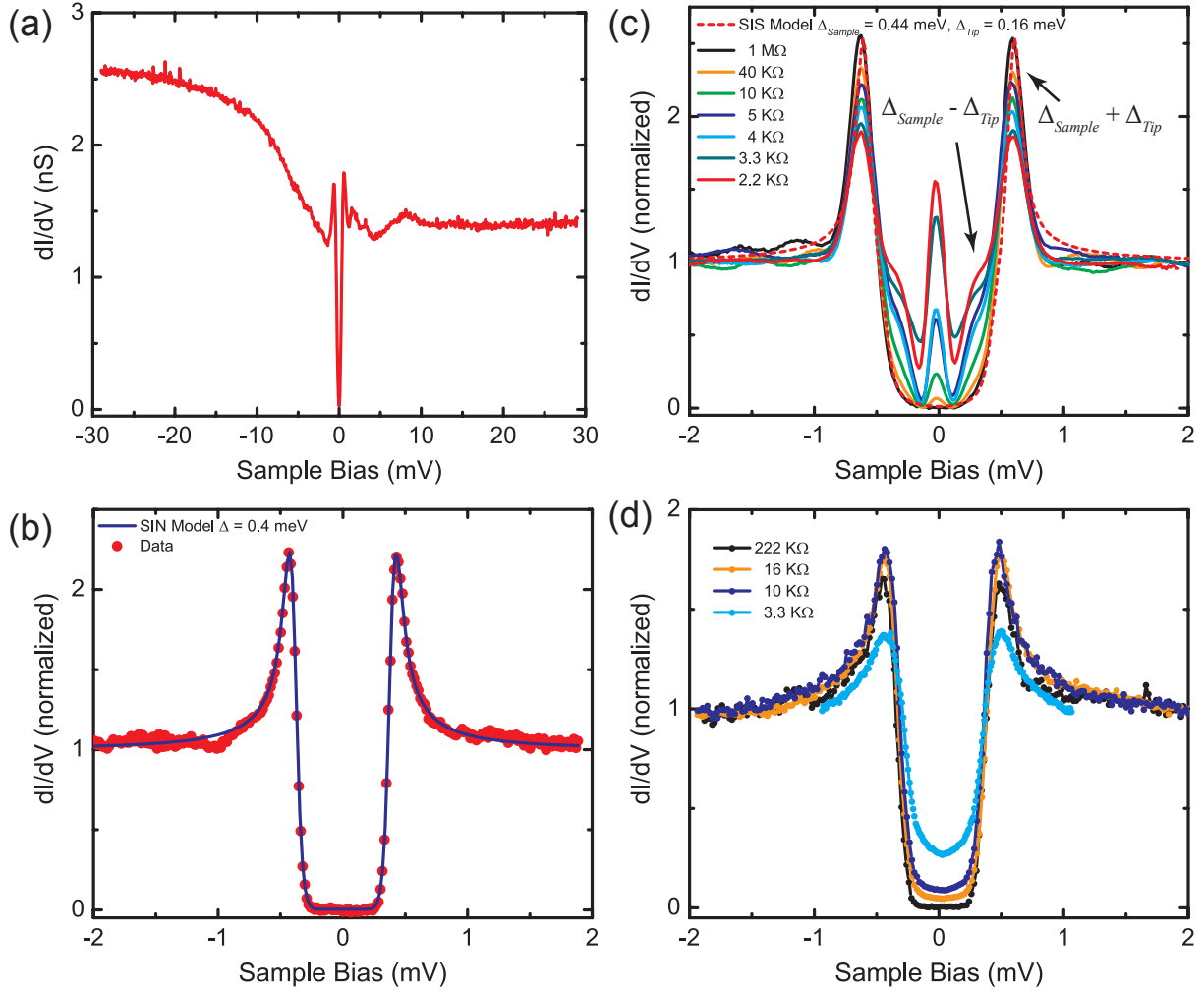

**Fig. 2**. Tunneling spectroscopy of $Cu_{0.2}Bi_2Se_3$. (**a**) Tunnel spectrum covering ±30 mV showing the background and superconducting gap near zero bias. Tunnel parameters: $I$=70 pA, $V_B$=29 mV, $V_{mod}$=100 μV. (**b**) Low energy tunnel spectrum (red dots) showing a fully gapped $Cu_{0.2}Bi_2Se_3$ consistent with BCS s-wave pairing. A fit (blue line) to BCS Maki theory with $T_{eff}$ = 280 mK yeilds Δ =0.399±0.001 meV and ζ=0.0092±0.0002 [20]. Tunnel parameters: $I$=300 pA, $V_B$=2 mV, $V_{mod}$=15 μV. (**c**) Tunnel spectrum of $Cu_{0.2}Bi_2Se_3$ with a probe tip crashed into the sample as a function of junction impedance. At high impedance the spectrum is well fit (red dashed line) by a BCS SIS model with the sample $\Delta_{Sample}$=0.44± 0.01 meV and a probe tip gap $\Delta_{Tip}$=0.16±0.01 meV, and $T_{eff}$ = 390±100 mK [20]. At lower junction impedance the spectrum develops a peak at zero bias due to Josephson tunneling. Tunnel parameters: $V_B$=2 mV, $V_{mod}$=10 μV, $I$=2 nA to 900 nA. (**d**) Tunneling spectrum on a second cleavage of the sample in (a) with a new probe tip showing no evidence of zero bias states at low impedance with a normal probe tip. Tunnel parameters: $V_{mod}$=20 μV, $V_B$=1 mV to 4 mV, , $I$=18 nA to 300 nA.

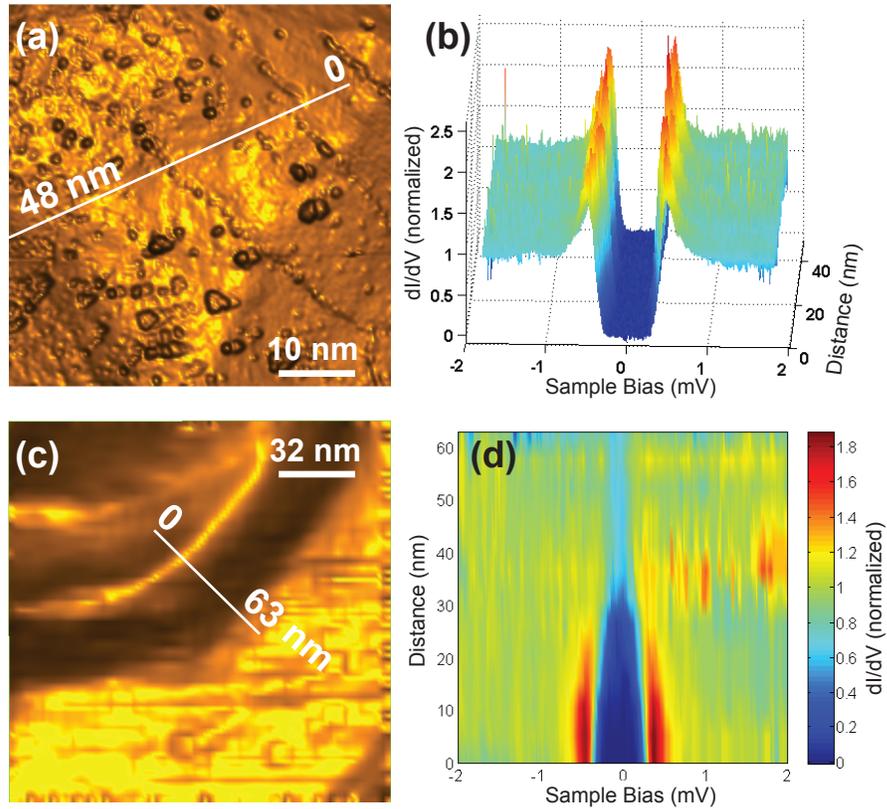

**Fig. 3**. Superconducting energy gap inhomogeniety in $Cu_{0.2}Bi_2Se_3$. (**a**) STM topographic image, 50 nm x 50 nm, of a superconducting terrace region. The white line indicates where the *dI/dV* measurements were made in (b). (**b**) *dI/dV* spectra from the region in (a) showing a relatively uniform energy gap across the terraces, grain boundaries, and Cu clusters in (a). The average gap from BSC fits is 0.41 ± 0.03 meV. Tunnel parameters: *I*=400 pA, $V_B$ =5 mV, $V_{mod}$ =20 µV. (**c**) STM topographic image, 160 nm x 160 nm, of a superconducting - normal interface region. The white line indicates where the *dI/dV* measurements were made in (d). (**d**) *dI/dV* spectral map along the line in (c) showing the gradual disappearance of the gap across the boundary. Each horizontal row is the *dI/dV* vs. $V_B$ spectra, where the *dI/dV* magnitude is given by the color intensity scale and the vertical axis is the distance along the line in (c). Tunnel parameters: *I*=200 pA, $V_B$ =5 mV, $V_{mod}$ =20 µV.

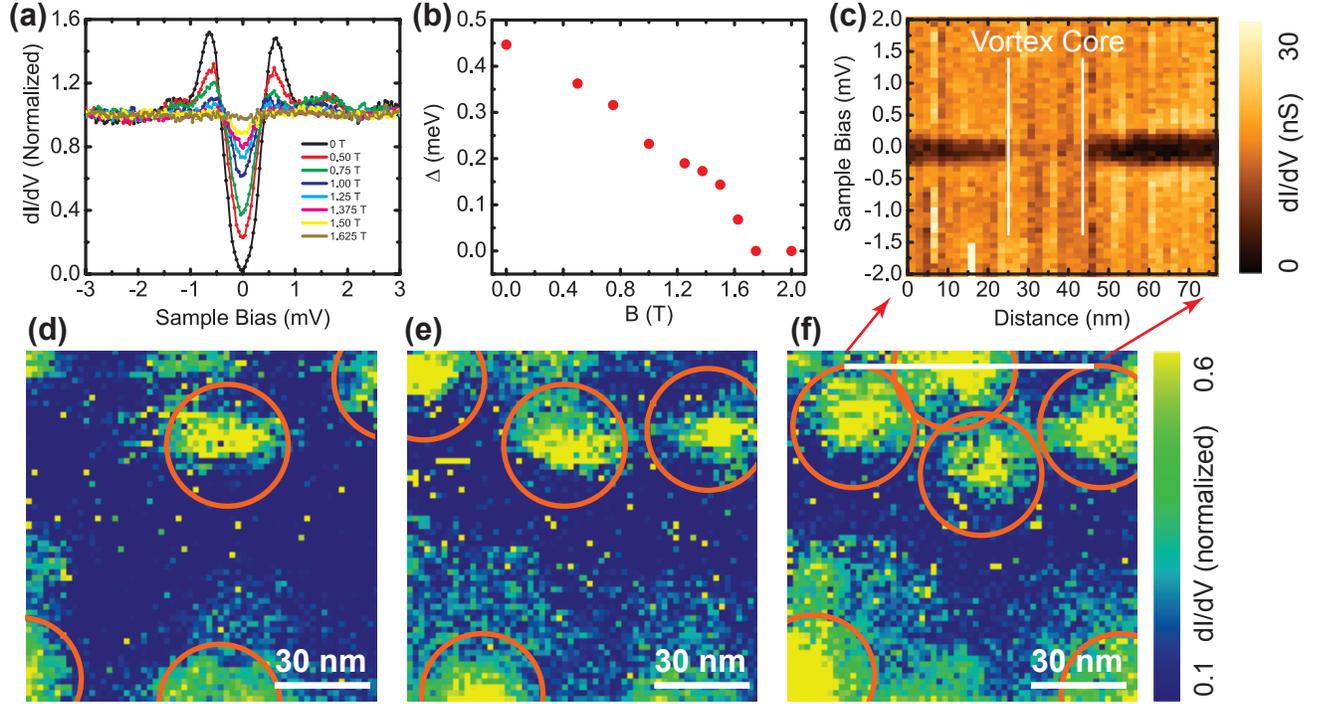

**Fig. 4**. Superconducting properties of $Cu_{0.2}Bi_2Se_3$ in an applied magnetic field. (**a**) Tunneling spectra as a function of an applied perpendicular magnetic field, $B$. The spectra show a transition to the normal state above $B$=1.7 T. The spectra are normalized to the spectra at high field ($B$=1.75-2 T). Tunnel parameters: $I$=80 pA, $V_B$=10 mV, $V_{mod}$=50 µV, $T_{eff}$=0.95 K. (**b**) The superconducting energy gap determined from fitting the spectra in (a) to the BCS density of states. The uncertainties given by one standard deviation representing one standard deviation in the gap energy determined from the fit to the BCS density of states are smaller than the symbol size [20]. (**c**) $dI/dV$ spectral map showing the disappearance of the superconducting gap in the center of the vortex core in the top of the image in (f). Each vertical line is a $dI/dV$ vs $V_B$ spectra, and the horizontal axis is the distance along the line in (f). The color scale is the $dI/dV$ intensity from 0 to 30 nS. (**d-f**) Density of states maps at the Fermi level, $dI/dV(V_B$=0 mV$) / dI/dV(V_B$=-0.5 mV$)$, 110 nm x 110 nm, showing the evolution of vortices (orange circles) as a function of applied magnetic field, (d) 0.5 T, (e) 0.75 T, and (f) 1.0 T. The white line in (f) indicates the path for the spectral map in (c). Tunnel parameters: $I$=200 pA, $V_B$=5 mV, $V_{mod}$=50 µV.